\documentclass[pdflatex,sn-mathphys-num]{sn-jnl}

\usepackage{graphicx}%
\usepackage{multirow}%
\usepackage{amsmath,amssymb,amsfonts}%
\usepackage{amsthm}%
\usepackage{mathrsfs}%
\usepackage[title]{appendix}%
\usepackage{xcolor}%
\usepackage{textcomp}%
\usepackage{manyfoot}%
\usepackage{booktabs}%
\usepackage{algorithm}%
\usepackage{algorithmicx}%
\usepackage{algpseudocode}%
\usepackage{listings}%
\usepackage{placeins}
\usepackage{chemformula}%

\begin{document}

\title[Article Title]{Quantum Computing in Corrosion Modeling: Bridging Research and Industry}


\author[1]{\fnm{Juan Manuel} \sur{Aguiar Hualde}} \email{juan-manuel.aguiar-hualde@capgemini.com}
\author[1]{\fnm{Marek} \sur{Kowalik}} \email{marek-jozef.kowalik@capgemini.com}
\author[1]{\fnm{Lian} \sur{Remme}} \email{lian.remme@capgemini.com}
\author[1]{\fnm{Franziska Elisabeth} \sur{Wolff}} \email{franziska-elisabeth.wolff@capgemini.com}
\author[1]{\fnm{Julian van} \sur{Velzen}} \email{julian.van.velzen@capgemini.com}
\author[2]{\fnm{Walden} \sur{Killick}} \email{walden.killick@cambridgeconsultants.com}
\author[3]{\fnm{Rene} \sur{B\"ottcher}} \email{rene.boettcher@airbus.com}
\author[3]{\fnm{Christian} \sur{Weimer}} \email{christian.weimer@airbus.com}
\author[3]{\fnm{Jasper} \sur{Krauser}} \email{jasper.krauser@airbus.com}
\author*[4]{\fnm{Emanuele} \sur{Marsili}}\email{emanuele.marsili@airbus.com}

\affil[1]{\orgname{Capgemini Quantum Lab}}
\affil[2]{\orgname{Cambridge Consultants}}
\affil[3]{\orgname{Airbus, Central Research \& Technology}, \orgaddress{\street{Willy-Messerschmitt-Stra{\ss}e 1}, \city{Taufkirchen} \postcode{82024},  \country{Germany}}}
\affil[4]{\orgname{Airbus, Central Research \& Technology}, \orgaddress{\street{Pegasus House Aerospace Ave}, \city{Bristol} \postcode{BS34 7PA},  \country{United Kingdom}}}

\newcommand{\ema}[1]{\textcolor{blue}{#1}}
\newcommand{\rene}[1]{\textcolor{red}{#1}}
\newcommand{\franzi}[1]{\textcolor{violet}{#1}}
\newcommand{\marek}[1]{\textcolor{orange}{#1}}
\newcommand{\jasper}[1]{\textcolor{green}{#1}}


\abstract{
Accurately modeling corrosion effects remains a significant challenge in ensuring the reliable prediction of product performance across various industries. This is especially critical for products exposed to harsh environmental conditions over their lifetime, such as those in the aerospace sector. In this work, we demonstrate a hybrid workflow that combines classical and quantum computing to simulate corrosion reactions using quantum algorithms. Our approach focuses on the initial step of the oxygen reduction reaction, a key trigger in the corrosion of typical aerospace aluminum alloys.

We explore, for the first time in this context, both noisy intermediate-scale and fault-tolerant quantum algorithms. We assess their performance in modeling the problem and conduct a detailed quantum resource estimation. Our results show that significant advances are required not only in quantum hardware but also in algorithms and error correction techniques to make quantum computation practically viable. Nonetheless, this work establishes a critical foundation for leveraging quantum computing in corrosion modeling and underscores its potential to address complex, business-relevant challenges in materials science.}


\maketitle

\section{Introduction} \label{sec1}
Corrosion is defined as the degradation of materials, particularly metals, due to chemical or electrochemical reactions with their environment~\cite{harsimran2021overview}, representing a significant challenge across industries, particularly aerospace, where it affects operational efficiency, vehicle lifespan, and maintenance planning and effort. In this sector, a wide variety of aluminum alloys (AA) are utilized, with the Cu-rich 2xxx series being one of the most commonly used due to their high specific strength~\cite{rambabu2017aluminium, jegdic2020corrosion}. However, the Cu-rich intermetallic phases (IMP) in these alloys make them highly susceptible to various forms of corrosion~\cite{abodi2012modeling}, including pitting corrosion, stress-corrosion cracking, and galvanic corrosion~\cite{frankel2008understanding}, which can compromise structural integrity and performance. Studies indicate that corrosion contributes significantly to maintenance costs and downtime in aerospace, with annual global costs exceeding \$2.5 trillion~\cite{koch2016international}.\\

Preventive measures and proactive decisions, such as scheduling maintenance or assessing corrosion stages, often rely on properties measured at the macroscopic level, such as chemical concentrations or the formation of byproducts. These observations are typically derived from experimental data and are frequently supported by computational models capable of accurately describing the entire corrosion process. A common approach to developing computational models for corrosion involves addressing its inherently multi-scale nature by decomposing the problem. At the atomic scale, the focus lies on capturing the fundamental chemical reactions occurring within the system, while the mesoscopic and macroscopic scales are typically modeled using finite volume (FVM) or finite element methods (FEM) to describe larger-scale behaviors and interactions~\cite{gunasegaram2014towards}. 

Over the years, FVM and FEM have been used to model corrosion in several systems, shedding light on corrosion progression, surface passivation mechanisms, and strategies for corrosion inhibition~\cite{xiao2011predictive, murer2012finite, derose2013aluminium, duddu2016extended, saeedikhani2020finite}. The accuracy of FVM and FEM depends on the precision with which the kinetic rates of derived chemical reactions are quantified at the alloy surface and within the electrolytes, given their coupled nature. Although some parameters can be reliably estimated using well-established experimental techniques, others are barely accessible and often require highly complex and expensive experimental equipment~\cite{wang2022local, snihirova2016corrosion}. This challenge is particularly evident for the oxygen reduction reaction (ORR, \ch{O_2 + 4 e^- -> 2 O^{2-}}).

\begin{figure}[h]
\centering
\includegraphics[width=0.9\textwidth]{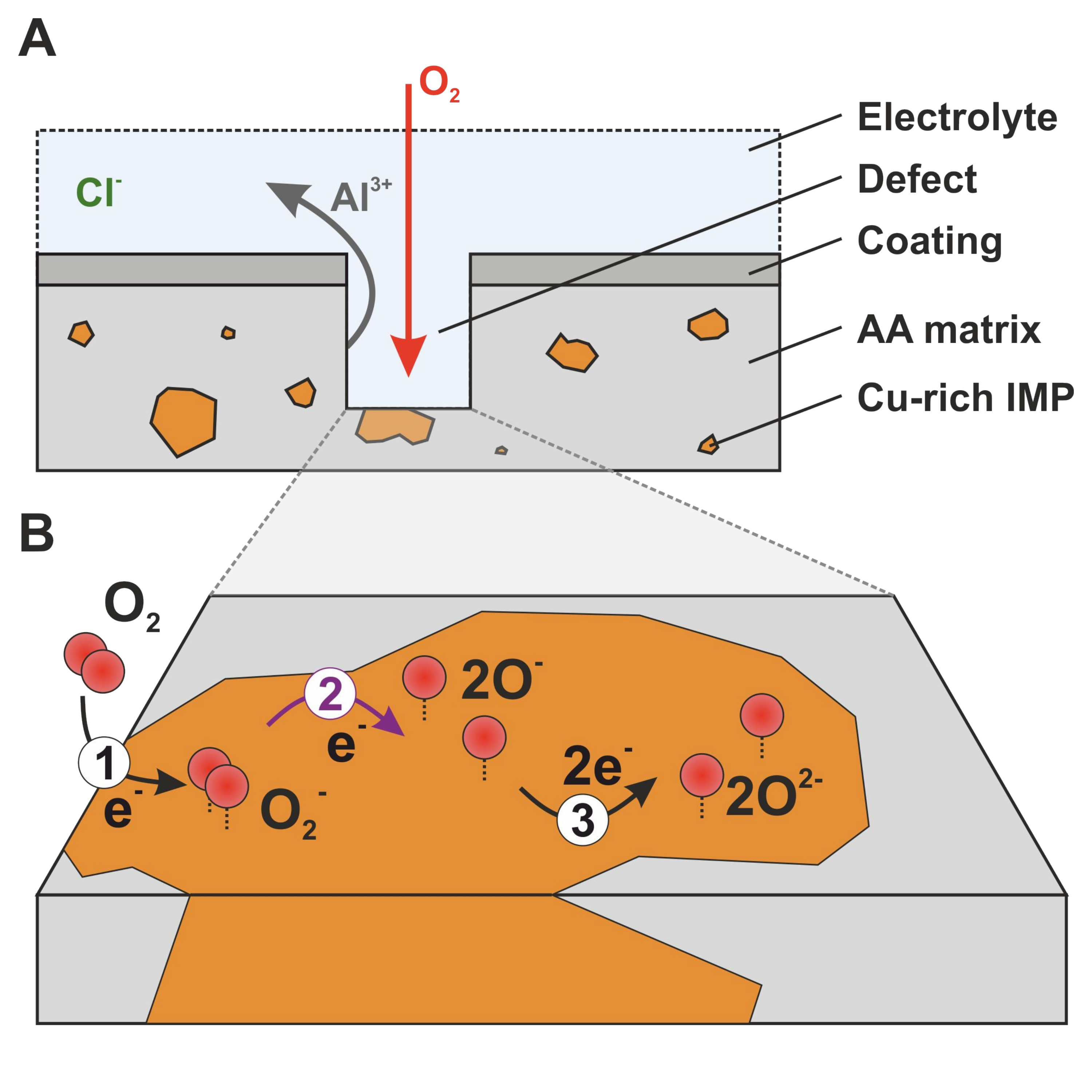}
    \caption{(A) Schematic cross section of a coated Cu-containing aluminum alloy (AA) with a U-shaped defect immersed in an electrolyte and (B) Cu-rich intermetallic particle (IMP) at the bottom of the defect with reaction pathway of the oxygen reduction reaction (ORR).} \label{fig1}
\end{figure}

In 2xxx alloys, the ORR plays a critical role as it is the cathodic half-reaction of the overall redox process. In case of the breakdown of any protective layer, such as coatings or natural oxide layers, his reaction is catalyzed by the exposed IMPs (Fig.~\ref{fig1}A, orange particles), which transform into nearly pure Cu-rich spots in the early stages of corrosion by releasing less noble elements first and acting as cathodic sites~\cite{kosari2020dealloying}. 
This accelerates the ORR and creates galvanic stress with the less noble surrounding aluminum matrix, which corrodes (Fig.~\ref{fig1}A, gray arrow)~\cite{boag2011corrosion}. 
The ORR is initiated once the oxygen dissolved in solution (Fig.~\ref{fig1}A, red arrow) undergoes reductive adsorption on the IMP surface (\ch{O_{2 \: (aq)} + e^- -> O^-_{2 \: (ad)}}, Fig.~\ref{fig1}B step 1), forming the superoxide ion which then undergoes to a reductive dissociation (\ch{O^-_{2 \: (a)} + e^- -> 2 O^-_{(a)}}, Fig.~\ref{fig1}B step 2). Subsequently, further reduction of the \ch{O^-} ions completes the ORR cycle with a total extraction of \ch{4 e-} from the metal surface (Fig.~\ref{fig1}B step 3). Although steps 1-3 take place at the IMP, the difference in the Nernst potential results in spontaneous electron transfer between the Cu-rich IMP and the aluminum matrix, which in turn is oxidized and releases dissolved aluminum ions into the electrolyte. The oxidation of the aluminum is the main driver of visible degradation of the alloy surface, \textit{e.g.}, pitting corrosion, which compromises its mechanical properties over time.\\

Given the multi-electron transfer pathways and the presence of reactive intermediates illustrated in Fig.~\ref{fig1}B,  it is not surprising that the reaction kinetics of the ORR is very challenging to evaluate experimentally. Fortunately, atomistic simulations offer a powerful computational toolbox to calculate kinetic rates \textit{in silico}, providing mechanistic insight and serving as a powerful tool to investigate how environmental conditions and alloy composition affect reactivity~\cite{exner2017kinetics}. In this context, a substantial body of research has focused on modeling the ORR in various metals and alloys using classical quantum chemistry~\cite{chu2017effects, liu2018oxygen, ke2019density, li2024computational}. More recently, hybrid approaches that combine quantum chemistry methods with both classical and quantum computing have been applied in material science~\cite{di2023platinum, alexeev2024quantum, hariharan2024modeling}, including studies that are relevant to corrosion, such as the dissociation of water on a magnesium surface~\cite{gujarati2023quantum}.

In this work, we present for the first time a hybrid quantum computing workflow to model the reductive dissociation of oxygen at the molecular level on a copper slab, simulating the processes occurring on Cu-rich aluminum alloys (AA). Specifically, in Sections~\ref{subsec2.1} and~\ref{subsec2.2}, we employ density functional theory (DFT), Hartree-Fock (HF), and post-HF methods to identify the critical geometries and the reaction pathways, determining the activation energies and kinetic rates while establishing a reference for strong correlation effects. In Section~\ref{subsec2.3}, we use these results to construct an embedding Hamiltonian for quantum simulations, where we employ the variational quantum eigensolver (VQE) and quantum phase estimation (QPE) algorithms to solve the electronic structure problem and benchmark their performance and runtime. Finally, we conclude by discussing the implications of these results for the feasibility of quantum approaches in modeling chemical processes of practical importance, such as corrosion.

\section{Results and Discussion} \label{sec2}

The kinetic rate of a chemical reaction ($k$) can be calculated using the standard transition state (TS) theory~\cite{eyring1935activated, bard1983electrochemical} via Arrhenius' law $k = A e^{-E_a/(RT)}$, where $A$ is the frequency factor, representing the frequency of collisions between reacting molecules, and $E_a$, that is the activation energy. $R$ is the universal gas constant and $T$ is the absolute temperature. $E_a$ corresponds to the minimum energy required for a reaction to occur, representing the energy barrier that the system must overcome along the reaction pathway. In this work, we focus on the path of the reductive dissociation reaction that takes place after the oxygen molecule is adsorbed on the surface of the alloy (Fig.~\ref{fig1}B step 2). This choice is motivated by two factors: (i) we anticipate this reaction to be the bottleneck of the overall oxygen reduction reaction (ORR), \textit{ i.e.}, the step with the highest energy barrier that governs the overall reaction rate, and (ii) we speculate that this pathway includes strong correlation effects due to the electron transfer that alter the electronic character of the ground state and therefore suitable to be studied using quantum computing approaches.

\subsection{The electronic Nature of the reductive Dissociation} \label{subsec2.1}

\begin{figure}
\centering
\includegraphics[width=0.9\textwidth]{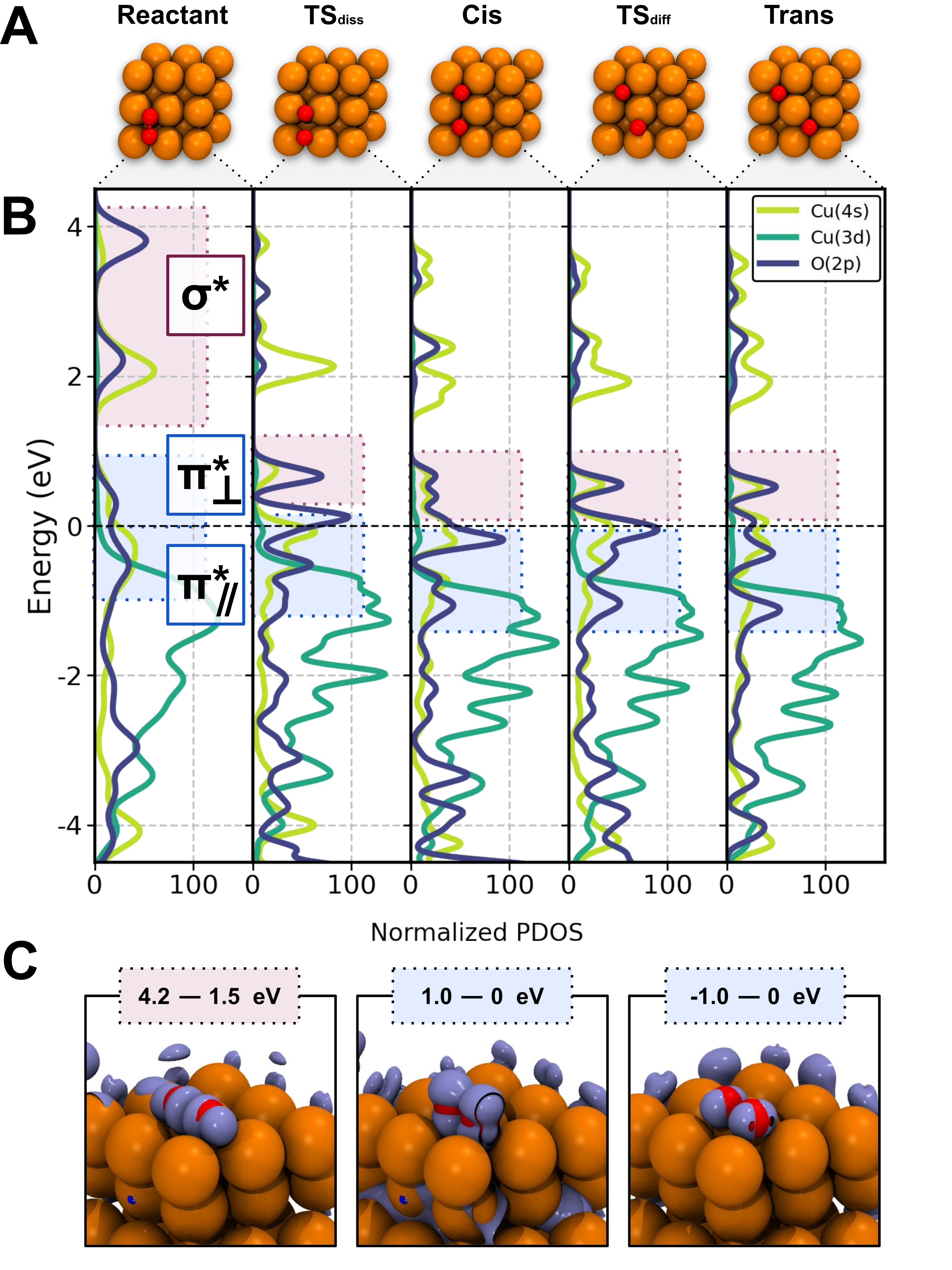}
\caption{(A) Top view of geometries identified along the reaction pathway: Reactant, oxygen dissociation (TS$_{\text{diss}}$), \textit{cis} configuration, oxygen diffusion (TS$_{\text{diff}}$) and \textit{trans} configuration. (B) PDOS for each geometry in A; we show those levels relevant for the oxidation: O(2p) in blue, Cu(3d) green and Cu(4s) light green. Regions associated to molecular orbitals $\sigma^*$ and $\pi^*$ (parallel and perpendiculars to the surface) are highlighted in light red and light blue respectively. (C) Graphical representation of the charge distribution integrated over $[4.2-1.5]$ eV, $[1.0-0]$ eV and $[-1.0-0]$ eV energy windows at the reactant geometries.} \label{fig2}
\end{figure}

To map the reductive dissociation reaction path, we use a simplified model representing Cu-rich IMPs. This model is described as a pure copper supercell, with periodic boundary conditions applied in the in-plane directions and a vacuum imposed in the direction perpendicular to the adsorption sites. Using this supercell, we optimized three critical geometries: The reactant configuration, where the molecular oxygen is adsorbed on the copper supercell, and the two possible product configurations (\textit{cis} and \textit{trans}, Fig.~\ref{fig2}A). 

Based on these three geometries, we trace the reaction profile by employing the nudged elastic band (NEB) with DFT, which generates a series of interpolated geometries between the initial and final states, including approximate TS geometries for each reaction step. In this case, we identify two transition states: the first one connects the reactant and the \textit{cis} configuration and is associated with the reductive dissociation reaction itself (TS$_\text{diss}$), while the second one connects the \textit{cis} and \textit{trans} configurations, corresponding to the diffusion of the two oxygen ions across the supercell slab (TS$_\text{diff}$). 

\subsubsection{Density of States at Reactant Geometry: First Electron Transfer} \label{subsubsec2.1.1}

Next, we focus on identifying the electronic nature of the five geometries. We begin by analyzing the projected density of states (PDOS) over three sets of atomic orbitals: oxygen 2p (O(2p)), copper 3d (Cu(3d)), and copper 4s (Cu(4s)), as shown in Fig.~\ref{fig2}B for each of the five geometries. Starting with the reactant, our attention is focused on the position of the electronic bands projected onto the O(2p) orbitals, specifically those with $\pi^*$(\ch{O-O}) and $\sigma^*$(\ch{O-O}) character. The $\pi^*$(\ch{O-O}) bands are located near the Fermi energy (Fig.~\ref{fig2}B region highlighted in light blue) and, compared to free molecular oxygen, are split into parallel and perpendicular components due to the interaction with the metal surface, resulting in the loss of degeneracy (see discussion for Fig. SI10 in supplementary information). The two components can be disentangled by plotting the integrated densities in two energy windows: The $[1, 0]$ eV window is associated with the $\pi^*_{\perp}$ orbital, while the $[-1, 0]$ eV range is characterized mostly by the $\pi^*_{\parallel}$ orbital (Fig.~\ref{fig2}C). In contrast, the $\sigma^*$(\ch{O-O}) band lay in the $[4.2 - 1.5]$ eV energy window (Fig.~\ref{fig2}C red) above the Fermi energy, indicating that it remains fully unoccupied.

A closer inspection shows that, already in the reactant geometries, the two oxygen atoms posses an excess of approximately one electron, as can be confirmed by looking at Fig.~\ref{fig3}A showing the net charge of the oxygen atoms. As mentioned above, the oxygen molecule has already accepted one electron during the adsorption process, resulting in the formation of a superoxide ion (\ch{O^-_{2}}). 

\subsubsection{Density of States along the Reaction Path: Second Electron Transfer} \label{subsubsec2.1.2}

Examining the $\pi^*$(\ch{O-O}) and $\sigma^*$(\ch{O-O}) bands at TS$_\text{diss}$, we observe that all of them shift towards lower energies, with the $\sigma^*$(\ch{O-O}) band approaching the Fermi energy level. This shift is a clear indication that the electronic character of the molecular system is changing (electron transfer occurs from the metal surface to the molecular oxygen). As the distance between the two oxygen atoms increases, the $\sigma^*$(\ch{O-O}) band becomes strongly stabilized, promoting the second electron transfer, resulting in the accumulation of approximately two excess electrons, one on each oxygen atom (see the charge at the \textit{cis} configuration in Fig.~\ref{fig3}A). The stabilization of the $\sigma^*$(\ch{O-O}) band further reduces the bonding energy of the superoxide ion, weakening the \ch{O-O} bond and favoring the dissociation process.


Finally, we conclude by observing that no additional electron transfer occurs once the \textit{cis} product is formed, as indicated by the minimal changes in the PDOS profile beyond this point. The second segment of the reaction path therefore represents a purely diffusive process of oxygen atoms across the copper lattice, with little to no impact on the electronic structure of the molecular system. 

\subsection{Energy Profiles along the Reaction Path} \label{subsec2.2}

\begin{figure}[h]
\centering
\includegraphics[width=0.95\textwidth]{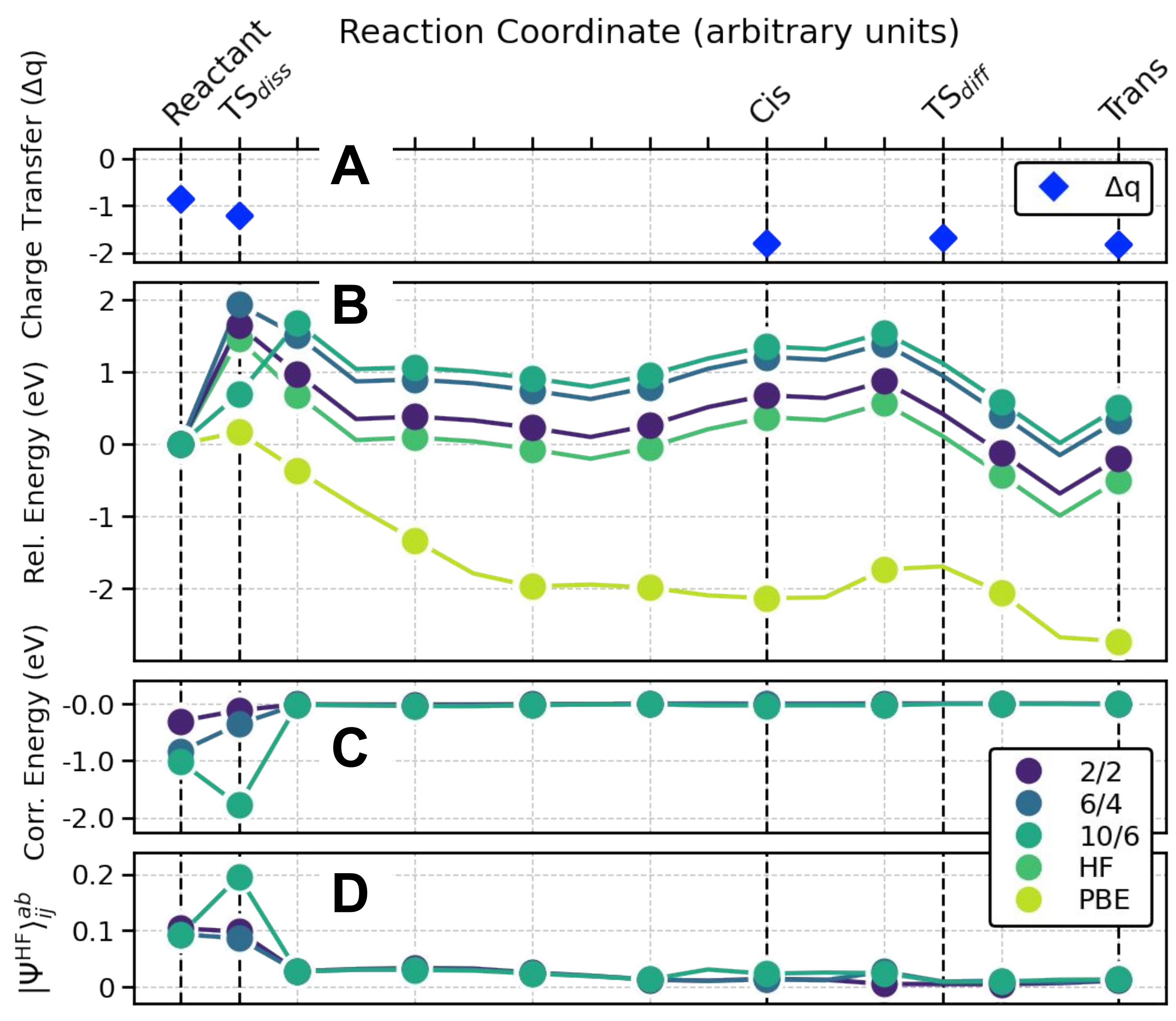}
\caption{Charge, energy and electronic correlation profiles along the reaction pathway: (A) Charge on each oxygen atom relative to the oxygen atom charge in O$_2$ using DFT (PBE) (B) Relative energy profiles calculated using DFT (PBE), HF, and FCI for active spaces (2/2) in purple, (6/4) in blue-green and (10/6) in green. (C) Correlation energy differences between FCI and HF along the reaction path. (D) Largest contribution of the doubly excited determinant appearing in the FCI wavefunction.} \label{fig3}
\end{figure}

The change on electronic character occurring at the TS$_\text{diss}$ geometry has significant implications for the energy profile along the reaction pathway. To understand this, we use the set of five geometries (Fig.~\ref{fig2}A), along with the interpolated geometries generated via the NEB technique using DFT, to compute with HF and post-HF methods the relative electronic energy along the reaction path (Fig.~\ref{fig3}B). We calculate the energy profile using DFT with the Perdew-Burke-Ernzerhof (PBE) exchange-correlation functional~\cite{perdew1996generalized} (Fig.~\ref{fig3}B yellow) and HF (Fig.~\ref{fig3}B light green).

HF predicts a significantly higher energy barrier than DFT and shows the energy basis not centered at the \textit{cis} and \textit{trans} products. However, these discrepancies are not surprising, as the geometries were optimized using PBE, which leads to suboptimal bond distances for HF. Moreover, the treatment of electronic interactions in DFT and HF differs significantly. In fact, HF is entirely incapable of accounting for both static correlation -- which occurs when multiple electronic configurations contribute to the ground state (multiconfigurational character) -- and dynamic correlation, related to orbital relaxation. In contrast, DFT can accurately describe dynamic correlation with a suitable functional~\cite{koch2015chemist, grimme2006semiempirical}.

\subsubsection{Static Electron Correlation along the Reaction Path} \label{subsubsec2.2.1}

To include static correlation in the HF wavefunction, a common approach is to employ configuration interaction (CI). In CI, additional electronic configurations are introduced by generating excitations from the reference HF wavefunction. Electrons are excited from occupied orbitals in the reference configuration to unoccupied ones, producing single, double, and higher excitations. These excited configurations can significantly contribute to the description of the electronic ground state, marking the ground state as multiconfigurational. To make this process computationally efficient, the excitations are restricted to a subset of orbitals called active space (AS). The AS defines the most relevant orbitals involved in the chemical reaction. Active space sizes are denoted with $n_e/n_o$, where $n_e$ is the number of active electrons and $n_o$ is the number of active orbitals. In this study, each AS is constructed using the Atomic Valence Active Space (AVAS) method~\cite{sayfutyarova2017automated}. AVAS projects the HF molecular orbitals (MOs) onto an atomic orbital (AO) subspace, localizes and computes the overlap between them, and selects the most important orbitals on the basis of this overlap. Due to the nature of the reductive dissociation reaction, we choose the AO subspace comprising the 2p orbitals of oxygen and the 3d orbitals of copper. As already shown in Fig.~\ref{fig2}B, this AO subspace is critical since it is directly involved in electron transfer.

With the AS defined, we proceeded to compute the energy using the full configuration interaction (FCI) method, where all possible excitations are generated within the AS. We perform this calculation for three different active spaces: (2/2) in purple, (6/4) in dark green, and (10/6) in green (section Active Space in Supplementary Information). The resulting energy profiles are displayed in Fig.~\ref{fig3}B. It is important to note that, in all three active spaces, we included only one virtual orbital -- the $\sigma^*(\ch{O-O})$, which is associated with the bond-breaking orbital in the reductive dissociation of \ch{O_2}. As a result, the maximum degree of excitation that can be generated from the HF configuration is two-electron excitations.\\

From the comparison of these three profiles with the HF energy profile, we observe a significant variation in both the energy barrier and the relative energies of the products in comparison to those of the reactants. In fact, for the two largest active spaces, (6/4) and (10/6), the thermodynamic driving force is even reversed, with the products being higher in energy than the reactants. This suggests that electron correlation plays an important role along the reaction pathway. To pinpoint where correlation effects are most significant, we plotted the energy difference between the FCI and HF calculations at each point along the reaction path (Fig.~\ref{fig3}C). The correlation energy is more negative for the reactants and the TS$_\text{diss}$, while it becomes negligible for the rest of the path. In other words, the initial section of the reaction path is characterized by an electronic ground state with a strong contribution from the excited configurations. As seen in Fig.~\ref{fig3}D, which shows the largest coefficient of the doubly excited state determinant in the FCI vector, the initial portion of the reaction path exhibits a strong contribution from doubly excited state determinants, responsible for the large correlation energy observed. 

The appearance of multiple excited state determinants in the ground state for the reactant and TS$_\text{diss}$ geometries suggests that the usage of quantum computing to model these reactions may be appropriate. With the AS already defined, we can easily construct an embedding Hamiltonian that explicitly describes the electrons within the reduced Fock space spanned by the AS orbitals~\cite{battaglia2024general}, treating the remaining electrons as an average, effective one-body interaction. The selected electronic Hamiltonian, specific to each AS and geometry, is then transformed using the Jordan-Wigner method to prepare it for quantum computation. 

\subsection{The Electronic Structure Problem in a Quantum Computer} \label{subsec2.3}

At this point, our discussion focuses on determining the static correlation contribution (Fig.~\ref{fig3}C) using a quantum computing framework, alongside an analysis of the associated resource requirements. Our goal is to provide a clear understanding of the resources needed to run both the VQE algorithm, designed for near-intermediate scale quantum (NISQ) computers, and a standard QPE algorithm~\cite{kitaev1995quantum, nielsen2000quantum}, suitable for fault-tolerant quantum computers (FTQC).

\subsubsection{Noisy intermediate-scale quantum Approach} \label{subsubsec2.3.1}
\begin{figure}
    \centering
    \includegraphics[width=1\textwidth]{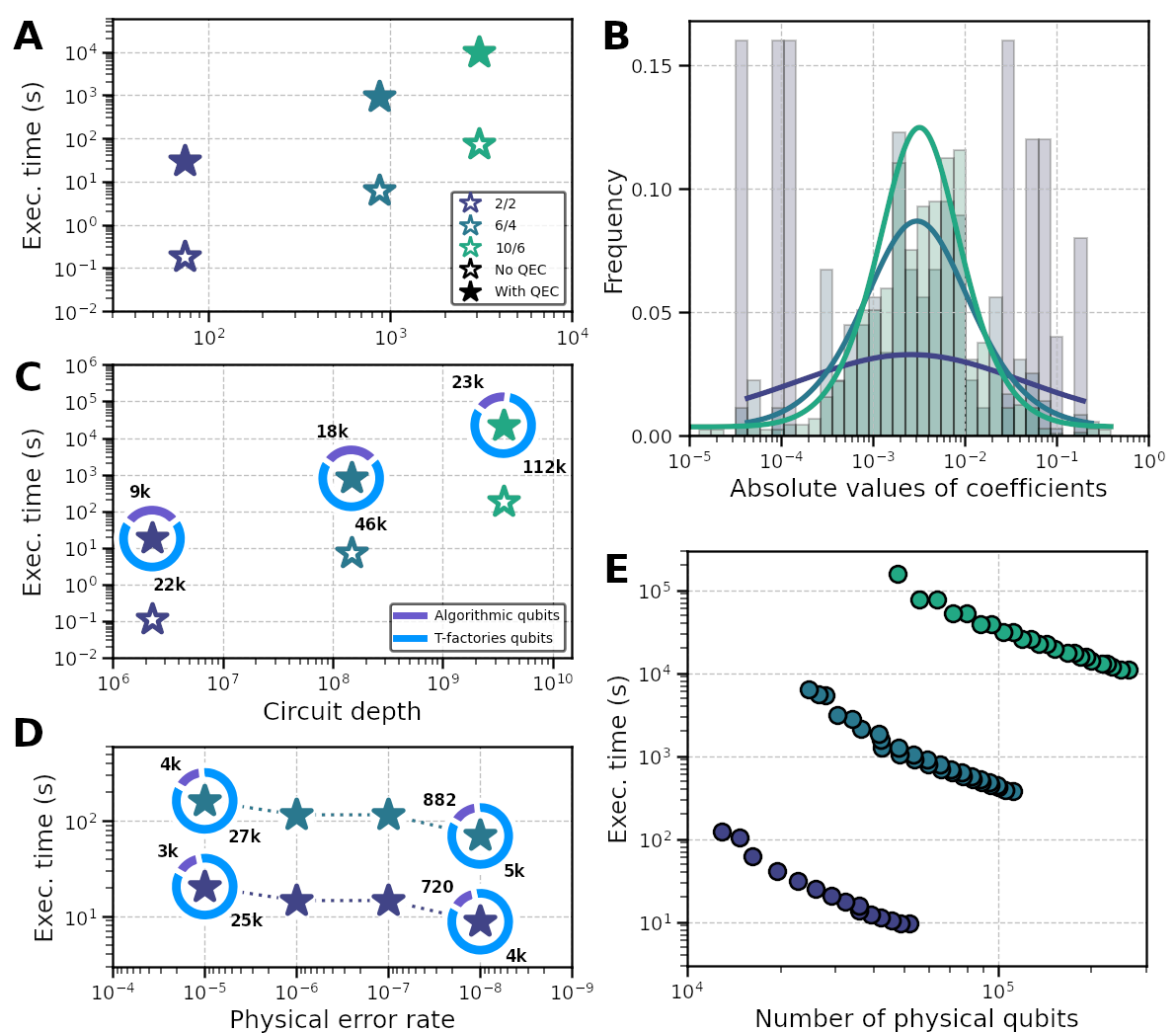}
    \caption{(A) Estimated execution runtime plotted against the circuit depths for the VQE calculations of the (2/2), (6/4), and (10/6) active space shown in purple, blue-green and green, respectively. Empty stars are the estimated denote calculations without considering the overhead of QEC, while full stars include the QEC based on the surface code. Panel (B) shows the histogram of the absolute values of coefficients with respect their magnitude appearing on the VQE ansatz. The distributions are fitted with Gaussian curves drawn with solid lines (scaled by arbitrary factor for better visualization). (C) Estimated execution runtime plotted against the circuit depths for the QPE. The pie chart shows the total number of physical qubits required, broken down in T-Factories qubits (bright blue) and the algorithmic qubits (purple). (D) Dependency of estimated execution runtime versus the error rate of the physical qubits. (E) Space-time frontier plot showing the dependency of the execution time against the total number of physical qubits.}
    \label{fig4}
\end{figure}

To present this comparison, we compiled the VQE quantum circuit (details of the calculation are provided in Methods) on a fake backend with the \texttt{Clifford}$+$\texttt{T} basis gate set and all-to-all connectivity. We compiled circuits for the three active spaces previously described to highlight the asymptotic scaling of the algorithm with increasing Hamiltonian embedding size. For each active space, we plot the quantum circuit depths and the runtime per each ansatz run with numerical values assigned to parameters (either for gradient evaluation or energy estimation for parameters), computed naively as the circuit depth multiplied by the ``wall clock time'' of the backend described in detail in Section~SI3. The results for runtime (in seconds) as a function of the circuit depth are shown as empty stars in Fig.~\ref{fig4}A. In addition, to account for quantum computer noise, we present also the estimated execution time of the quantum error-corrected (QEC) algorithm using the surface code as full stars. 

Analyzing Fig.~\ref{fig4}A, we observe that even with a modest active space such as (10/6), the execution time quickly becomes prohibitive, exceeding one minute even without considering error correction. The primary cost driver for the VQE algorithm is the increasing number of shots required to accurately sample the terms of the qubit-mapped Hamiltonian. To achieve a desired precision in the energy measurement (see Section~SI3), the number of shots must be proportional to the absolute values of the coefficients in the qubit-mapped Hamiltonian. Larger coefficients demand more shots to ensure a bounded variance in the Hamiltonian expectation value. Fig.~\ref{fig4}B shows the distribution of these coefficients together with their Gaussian fit. Interestingly, the standard deviation of the distribution systematically decreases as the active space size increases. This suggests that while the number of Hamiltonian terms grows polynomially with system size, the number of terms with large coefficients -- which primarily determine the required shot count -- does not grow as rapidly.

Despite this mitigating factor, we anticipate that VQE will struggle with larger system sizes, particularly when quantum error correction (QEC) is necessary for circuits with depths exceeding $10^3$. Even under optimistic assumption for the physical qubit error rate (here we consider $10^{-4}$), the runtime for the (10/6) active space surpasses one hour, which is significantly longer than the millisecond-level runtime achievable via exact diagonalization on classical computers.

\subsubsection{Fault Tolerant Approach} \label{subsubsec2.3.2}
We now turn our attention to the estimated execution runtime of the QPE calculation, shown in Fig.~\ref{fig4}C. As expected, the circuit depth is four to six orders of magnitude larger for the active spaces considered here. However, the estimated runtime for QPE, both with and without QEC, remains similar. This is because the favorable scaling of QPE is offset by the increasing number of ancilla qubits required to achieve the desired accuracy -- rising from 10 for the smaller active space to 15 for the larger -- which introduces an exponential cost in the runtime estimates. For instance, the estimated runtime for the (10/6) active space is approximately 2 hours when accounting for the overhead of the surface code, a duration significantly longer than what is achievable with classical methods. Moreover, since the QPE algorithm provides only a polynomial advantage with respect to classical methods -- due to the exponential reduction of the overlap between the HF wavefunction and the true ground state~\cite{lee2023evaluating} -- QPE is unlikely to offer significant benefits in solving the electronic structure problem, at least in the current premises.\\

To hypothesize about future scenarios in which QPE with QEC might be practical, let us consider quantum hardware with specifications surpassing those used so far. This scenario is illustrated in Fig.~\ref{fig4}D, where the runtime and the number of physical qubits required are plotted against the physical qubit error rate, ranging from $10^{-5}$ to $10^{-8}$. For the (6/4) active space, under such optimistic -- and perhaps improbable -- engineering advancements, the number of qubits required decrease drastically, dropping from 31 thousand for a $10^{-5}$ error rate to less than 6 thousand for a $10^{-8}$ physical error rate. On the other hand, the QPE execution runtime, which is closely tied to the logical cycle of the surface code, does not change substantially.

Finally, the execution time can also be reduced by adjusting the trade-off between the number of physical qubits and runtime, as allowed by the surface code. Intuitively, the space-time volume behaves like an incompressible fluid: increasing the number of qubits can reduce the execution time. Here, we examine whether a significantly larger quantum computer could substantially decrease the runtime, making QPE more efficient. However, as shown in Fig.~\ref{fig4}E, which presents the frontier plot for all three active spaces, while the runtime can be reduced by more than an order of magnitude, it remains insufficient. Even with this trade-off, the (10/6) active space runtime does not approach the one-hour mark. 

From this analysis we conclude that the usage of QPE is not limited only by the hardware available but it maybe still impractical without improvements to both algorithms and QEC techniques. Some of these improvements are already available today, including qubitization~\cite{low2019hamiltonian}, quantum walk operators~\cite{berry2018improved, poulin2018quantum}, approximation of the electron repulsion tensor~\cite{rocca2024reducing}, and more efficient error correction schemes~\cite{bravyi2024high}. Still, these advances may not be sufficient to make QPE competitive. In light of these results, it is clear that in addition to hardware improvements, algorithmic innovations must likewise be developed for quantum computers to be useful for industry-relevant applications.\\

Finally, as also highlighted by others~\cite{beverland2022assessing}, the number of physical qubits required (Fig.~\ref{fig4}C) is also concerning. The $\sim$135k qubits required to run QPE for the (10/6) active space -- primarily driven by the demands of the T factories -- pose a significant bottleneck. For context, Quantinuum anticipates building its Apollo chip by 2029 with ``only'' 1,000 physical qubits and a two-qubit error rate of $10^{-4}$, consistent with the error rate assumptions in our analysis. The fact that quantum computing, at least for chemistry applications, is a distant prospect is perhaps not surprising for many. However, our analysis underscores just how far we are from realizing any tangible industrial benefits from quantum computing, especially given that much larger Hamiltonians will need to be addressed before meaningful advantages can be achieved.

\section{Conclusion} \label{sec3}

To the best of the authors' knowledge, this study represents the first modeling of corrosion processes using a hybrid workflow that integrates both NISQ-oriented algorithms and fault-tolerant quantum computing approaches. We focus on one of the initial steps of the corrosion process: the ORR on a copper surface, a reaction that characterizes the corrosion of Cu-rich AA and that is of paramount importance for the oxidation of the aluminum substrate. 

The present study was divided into two critical steps. The first step involved assessing the ORR rate, which is essential for constructing macroscopic models of the reaction. This was achieved using classical quantum chemistry methods, including DFT, HF, and post-HF methods. This analysis revealed and quantified the activation energy barrier for the reaction while also highlighting the emergence of a strongly correlated ground state caused by a change in the electronic structure's nature. This finding is particularly significant because strongly correlated states are notoriously challenging for conventional classical methods, whereas quantum computers could offer a distinct advantage in handling them.\\

Armed with these insights, we turn our attention on how solving the electronic structure problem on a quantum computer. Specifically, we analyze the resource estimate to run the VQE and QPE quantum computing algorithms for the active space in hand. We find out that, for the small Hamiltonians considered, the runtime for VQE and QPE are not too far off. However, both cases we obtain runtime far higher than what classical methods can achieve already today.   

Unsurprisingly, the estimates for VQE with QEC reach a few minutes per iteration, due to surface code overhead and significant number of shots. VQE may requires between thousands and millions of circuit runs, depending on the number of parameters in the ansatz, that is why the small runtime of single experiment execution is crucial. 

On the other hand, while QPE is expected to scale more favorably, it suffers from the considerable overhead associated with QEC, imposing substantial penalties on both qubit count and runtime. Even under optimistic assumptions regarding physical qubit characteristics, the resource demands for QPE remain far from competitive with classical methods for the active space sizes explored here -- and likely for larger active spaces as well. Furthermore, even with significant improvements in physical qubit error rates, while the required number of qubits decreases significantly, the same cannot be said for execution runtime. 

In light of these findings, it has been concluded that improvements in physical qubit hardware alone are insufficient to make quantum computing viable, at least for finding the ground state of molecular systems. Instead, substantial advancements in algorithm design and QEC procedures are essential to make quantum computing useful in the domain of material science.\\


\section{Methods} \label{sec4}

\textbf{Quantum Chemistry Electronic Structure}. All DFT calculations (SCF and NEB calculations) were performed using the Quantum ESPRESSO 7.3~\cite{giannozzi2009quantum, giannozzi2017advanced, giannozzi2020quantum}. The PBE exchange-correlation functional~\cite{perdew1996generalized} were utilized within the generalized gradient approximation (GGA). We used ultra-soft pseudopotentials for copper and oxygen to efficiently represent electron-ion interactions while maintaining computational efficiency. 

The system was modeled as a three-layer slab of copper with a $3 \times 3$ arrangement of atoms in each layer, exposing a surface of the crystal plane (001). The slab was constructed with a 30 \AA~vacuum along the z axis with a \ch{O2} molecule placed on the surface. The molecular orbitals occupations were treated with Gaussian smearing using a smearing width of 0.01 Ry to facilitate convergence of metallic states. The electronic self-consistency was controlled by a local Thomas-Fermi mixing mode. A plane-wave kinetic energy cutoff of 50 Ry and a charge density cutoff of 200 Ry were applied based on convergence tests and the Brillouin zone was sampled at the Gamma point.

The structural relaxations, including the optimization of the reactants and the product as well as the NEB simulations, were performed until the forces in each atom were below 0.01 eV/\AA, with the electronic convergence threshold set to $10^{-8}$eV.

For electronic structure analysis, the projected density of states (PDOS) was calculated by projecting the Kohn-Sham states onto atomic orbitals, using a Gaussian smearing width of 0.02 Ry. The Fermi energy was set at zero for convenience in plotting PDOS data.\\

The HF wavefunction was obtained using PySCF 2.6.2, employing Gaussian-type orbitals as the basis set, with basis functions having exponential coefficients smaller than 0.1 discarded, along with an auxiliary basis set for density fitting. The slab structure optimized with DFT was modified by removing the bottom layer of copper atoms, resulting in a two-layer copper slab. To ensure consistency, the same kinetic energy cutoff and periodic boundary conditions used in the DFT calculations were applied to the HF calculations. The HF molecular orbitals (MOs) were subsequently utilized for correlated (post-HF) calculations and to construct the embedding Hamiltonian, which incorporates effective one- and two-body interactions~\cite{battaglia2024general}. The embedded Hamiltonian expressed in the second quantization form was transformed into qubit operator by Jordan-Wigner mapping~\cite{Jordan1928}.\\

\textbf{Quantum Computing}. The VQE is a hybrid quantum-classical algorithm for finding the energies of ground or excited states of a qubit-mapped Hamiltonian~\cite{peruzzo2014variational}. In this work, the method for solving the ground state estimation problem was applied. This approach is based on the Ritz variational principle~\cite{cerezo2021variational}, where by expressing the simulated system state through the parametrized quantum state $\Psi(\theta)$ on a quantum circuit, the lowest energy configuration is sought through tuning the parameters $\theta$ towards the ground state energy $E_0$:
$$
\min_{\theta}\left\langle \Psi(\theta) \right| \hat{H} \left| \Psi(\theta)\right\rangle \rightarrow E_0
$$
using the optimizer on classical computational backends to update the parameters and quantum backends to evaluate the gradients and energy for given set of parameters

The VQE routine was implemented with Qiskit 1.2~\cite{javadi2024quantum}. Number of shots was set up to bound standard deviation of estimated energy with threshold inspired by chemical accuracy $\epsilon=1.6$ mHa. Detailed analysis can be found in Section~SI2. The unitary coupled cluster with singles and doubles (UCCSD) ansatz was used, based on the selected active space and the initial state based on the HF wavefunction. UCCSD yields deep circuits in comparison to other methods (\textit{e.g.}, hardware efficient ansatz), but is chemistry-informed and has fewer parameters because the terms exponentiated in the ansatz are grouped sharing the same parameter. Detailed comparison of the depths and number of parameters per different ansaetzes in Section~SI2. 
The energy was measured with a qubit-mapped effective Hamiltonian of reactant in second quantized form. As an optimizer SLSQP~\cite{virtanen2020scipy} was used, due to its good empirical performance for VQE~\cite{morita2024simulator, powers2023using}. 
Additionally, Adapt-VQE, truncation of terms in UCCSD, and Clifford parameters warm start were tested with no improvements in accuracy and number of iterations was observed. All resource estimation results were generated for a generic implementation with built-in Qiskit 1.2 transpilation options with \texttt{optimization\_level}=1 and \texttt{ElidePermutations} pass included. The VQE converged towards analytically calculated ground state energies for three checked active space sizes: (2/2), (6/4) and (10/6). The results are shown in Section~SI2 at Figure~SI4.\\

The QPE algorithm was implemented using Qiskit 1.2~\cite{javadi2024quantum}. The Hamiltonian was scaled by a factor $\epsilon$ to ensure that the corresponding phase readout, $\phi_i \equiv 2\pi \lambda_i$ with $\{\lambda_i\}$ the eigenspectrum of the Hamiltonian, lies within the interval $\in (0, 2\pi]$ (see Section~SI1). Specifically, $\epsilon = 0.8$ a.u. and $\epsilon = 3.91$ a.u. were used for the embedding Hamiltonian in the (2/2) and (6/4) active spaces, respectively. These values were selected somewhat arbitrarily to be slightly higher than the modulus of the largest eigenvalue of the two operators. It is worth noting that this is not the most efficient method to extract the phase after running the QPE algorithm, as the resulting phases may not correspond to the closest discrete binary representation. However, we believe that this approach is valid for practical applications of QPE, where the eigenspectrum of the Hamiltonian can only be inferred approximately, and the full spectrum is not known. The full eigenspectra of the (2/2) active space is shown in Fig.~SI1.

The exponentiated Hamiltonian was approximated using the Lie-Trotter formula, with the accuracy of the approximation monotonically improving as the number of Trotter repetitions increased. We determined that 3 and 1 Trotter steps, along with 10 and 13 ancilla qubits, were sufficient to achieve chemical accuracy for the (2/2) and (6/4) active spaces, respectively (see Fig.~SI2 and Fig.~SI3). 

For both VQE and QPE calculation we used the quantum resource estimation tool provided by Microsoft Azure Quantum~\cite{beverland2022assessing, prateek2023quantum} to estimate the number of physical qubits and runtime of the algorithm. Further details about the quantum error correction parameters used are provided in Section~SI3. 

\backmatter

\bmhead{Supplementary information}
Supporting Information is available free of charge.

\bmhead{Acknowledgements}
We acknowledge the valuable insights provided by James Cruise from Cambridge Consultants during discussions on quantum error correction analysis. We also extend our gratitude to the Surface and Technology team at Airbus AIRTeC (Bristol, United Kingdom), particularly Jeremy Bradley (Senior Expert in Surface Protection and Corrosion Prevention), for their support and contributions to this project. Finally, we are deeply grateful to Airbus and Capgemini for fostering an excellent work environment focused on innovation and technological advancement.

\section*{Declarations}

\bmhead{Competing interests} 
The authors declare that they have no competing interests. 

\bmhead{Data availability} 
The authors declare that the data supporting the findings of this study are available in the main article and the Supplementary Information. Additional data are available from the corresponding author upon request. 

\bmhead{Author contributions} 
E.M. and R.B. conceived the project. E.M. and J.M.A.H. designed the classical chemistry calculations and analyzed the data. E.M. developed the embedding Hamiltonian formulation and performed the QPE calculations. M.K. and L.R. performed the VQE calculations. E.M. and W.K. performed the quantum resource estimation. All authors contributed to writing and reviewing the manuscript.

\bibliography{sn-article}

\end{document}


\title[Article Title]{Quantum Computing in Corrosion Modeling: Bridging Research and Industry}

\author[1]{\fnm{Juan Manuel} \sur{Aguiar Hualde}} \email{juan-manuel.aguiar-hualde@capgemini.com}
\author[1]{\fnm{Marek} \sur{Kowalik}} \email{marek-jozef.kowalik@capgemini.com}
\author[1]{\fnm{Lian} \sur{Remme}} \email{lian.remme@capgemini.com}
\author[1]{\fnm{Franziska Elisabeth} \sur{Wolff}} \email{franziska-elisabeth.wolff@capgemini.com}
\author[1]{\fnm{Julian van} \sur{Velzen}} \email{julian.van.velzen@capgemini.com}
\author[2]{\fnm{Walden} \sur{Killick}} \email{walden.killick@cambridgeconsultants.com}
\author[3]{\fnm{Rene} \sur{B\"ottcher}} \email{rene.boettcher@airbus.com}
\author[3]{\fnm{Christian} \sur{Weimer}} \email{christian.weimer@airbus.com}
\author[3]{\fnm{Jasper} \sur{Krauser}} \email{jasper.krauser@airbus.com}
\author*[4]{\fnm{Emanuele} \sur{Marsili}}\email{emanuele.marsili@airbus.com}

\affil[1]{\orgname{Capgemini Quantum Lab}}
\affil[2]{\orgname{Cambridge Consultants}}
\affil[3]{\orgname{Airbus, Central Research \& Technology}, \orgaddress{\street{Willy-Messerschmitt-Stra{\ss}e 1}, \city{Taufkirchen} \postcode{82024},  \country{Germany}}}
\affil[4]{\orgname{Airbus, Central Research \& Technology}, \orgaddress{\street{Pegasus House Aerospace Ave}, \city{Bristol} \postcode{BS34 7PA},  \country{United Kingdom}}}

\renewcommand{\figurename}{Fig. SI}

\maketitle

\section{Convergence of quantum phase estimation} \label{SI: sec1}
The first detail to consider when implementing the quantum phase estimation (QPE) algorithm is the bound of the eigenspectra of the Hamiltonian. Our goal is to map the real eigenvalues, ${\lambda_i}$, to points within the interval $(0, 1]$. This ensures that the corresponding phase readouts, $\phi_i \equiv 2\pi \lambda_i$, lie within the interval $(0, 2\pi]$. Such a procedure prevents the misassignment of eigenvalues caused by phases that wrap around the $2\pi$ range.

For instance, if the smallest eigenvalue of the Hamiltonian is equal to the negative of the largest eigenvalue, it becomes impossible to distinguish these two eigenvalues, as they are mapped to the same point in the interval. To resolve this, we can scale the eigenspectrum of the Hamiltonian by multiplying the operator by a constant factor $\epsilon$ such that $\lambda_i \epsilon \in (0, 1] \hspace{0.2cm} \forall \lambda_i$.

The Hamiltonian's exact eigenspectrum can be obtained by diagonalizing the Hamiltonian. Although this implies solving the problem classically (which defeats the purpose of using the QPE algorithm), it allows us to extract valuable information, such as the configurations that contribute most to the eigenvectors, including the ground state, and the coefficients associated with each basis state.

\begin{figure}[h]
\centering
\includegraphics[width=1\textwidth]{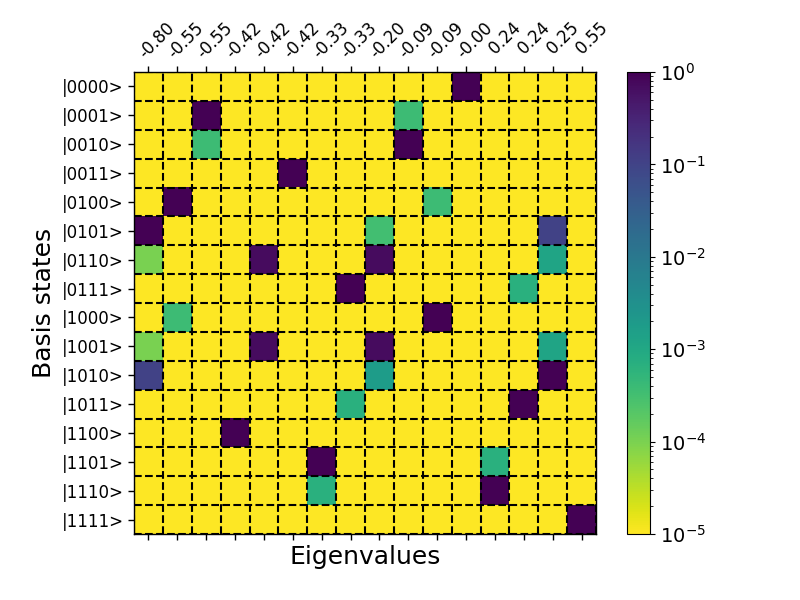}
\caption{Eigendecomposition of the Hamiltonian diagonalized over the complete Fock space comprising four fermionic states. The eigenvalues are plotted alongside the corresponding eigenvectors, which represent occupation number basis states ranging from the empty state $|0000\rangle$ to the fully occupied state $|1111\rangle$.} \label{fig hamiltonian 2-2}
\end{figure}

In Fig.~SI~\ref{fig hamiltonian 2-2}, we highlight the eigendecomposition of the full Hamiltonian constructed from the active space (2/2), diagonalized over the entire Fock space composed of four fermionic states. The basis states span all possible occupation number vectors, ranging from the empty state $|0000\rangle$ to the fully occupied state $|1111\rangle$. The Hartree-Fock (HF) state corresponds to the $|\text{HF}\rangle \equiv |0101\rangle$ basis state, which can be expressed more explicitly using energy and spin labels for the spin orbitals as $| 0_{1,\beta} 1_{0,\beta} 0_{1,\alpha} 1_{0,\alpha} \rangle$, with the two electrons populating the two lowest energy spin-orbitals.

For the ground state, which corresponds to the lowest eigenvalue (approximately $0.8$ a.u., represented by the leftmost eigenvector), the main contribution comes from the HF state. However, there is a non-negligible contribution from the doubly excited determinant $|1010\rangle$, which remains a pure single-spin state configuration, as expected, but with two electrons promoted to the $\sigma^*$(\ch{O-O}) orbital. Smaller contributions to the ground state are associated with singly excited determinants $|1001\rangle$ and $|0110\rangle$. Notably, the triplet electronic states, determined by the determinants $\{ |1100\rangle$, $\frac{1}{\sqrt{2}}(|0110\rangle - |1001\rangle)$, $|0011\rangle \}$, lie higher in energy and are thus considered excited states of the electronic Hamiltonian.

The second key component of the QPE algorithm is the trotterized form of the exponentiated electronic Hamiltonian, $e^{i \mathbf{H} \epsilon}$, where $\epsilon$ is the chosen scaling factor. To obtain the trotterized version of the Hamiltonian, we use the Lie-Trotter formula:
\begin{equation} 
    e^{i \epsilon \sum_k h_k} \approx \left( \prod_k e^{i h_k \epsilon / n} \right)^n \,, 
\end{equation} \label{eq: lie-trotter}
where $h_k$ are the terms comprising the electronic Hamiltonian, and $n$ is the number of Trotter steps. Equation~\eqref{eq: lie-trotter} becomes exact in the limit $n \rightarrow \infty$.

\begin{figure}[h]
\centering
\includegraphics[width=1\textwidth]{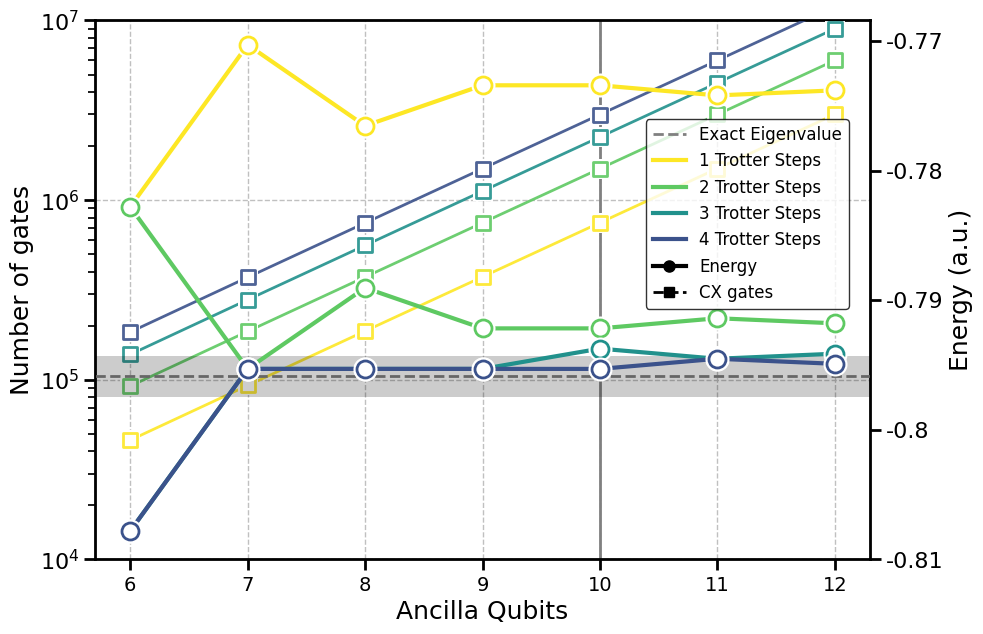}
\caption{Convergence results for the (2/2) Hamiltonian. The x-axis represents the number of ancilla qubits used to encode the binary representation of the phase. On the right y-axis, we plot the energy of the most-likely eigenvalue (empty circles), while the left y-axis shows the number of CNOT gates required for the calculation (empty squares). Lines connecting the circles and squares are included as visual guides. The results are computed using a varying number of Trotter steps, $n$, ranging from 1 (yellow) to 4 (purple). The exact eigenvalue, obtained by diagonalizing the Hamiltonian, is indicated by the horizontal dashed line. The gray-shaded area marks the chemical accuracy region, defined as $\lambda_{gs} \pm 1$ kcal/mol, where $\lambda_{gs}$ represents the ground-state eigenvalue. Additionally, the horizontal gray line shows the number of ancilla qubits required to achieve a phase representation precision of at least chemical accuracy.} \label{fig qpe convergence 2-2}
\end{figure}

\begin{figure}[h]
\centering
\includegraphics[width=1\textwidth]{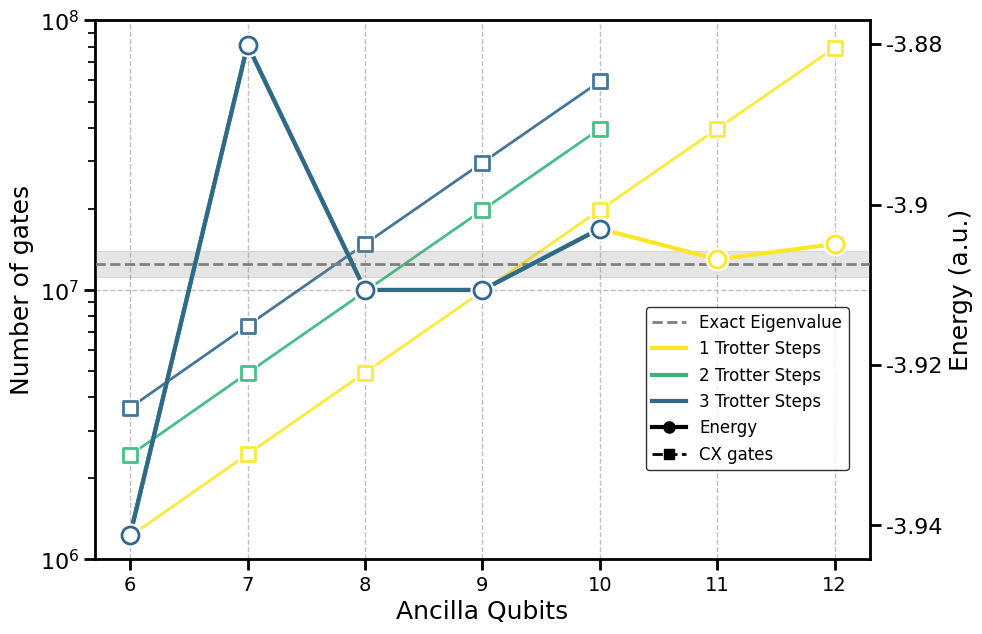}
\caption{Convergence results for the (6/4) Hamiltonian. The x-axis represents the number of ancilla qubits used to encode the binary representation of the phase. On the right y-axis, we plot the energy of the most-likely eigenvalue (empty circles), while the left y-axis shows the number of CNOT gates required for the calculation (empty squares). Lines connecting the circles and squares are included as visual guides. The results are computed using a varying number of Trotter steps, $n$, ranging from 1 (yellow) to 4 (purple). The exact eigenvalue, obtained by diagonalizing the Hamiltonian, is indicated by the horizontal dashed line. The gray-shaded area marks the chemical accuracy region, defined as $\lambda_{gs} \pm 1$ kcal/mol, where $\lambda_{gs}$ represents the ground-state eigenvalue.} \label{fig qpe convergence 6-4}
\end{figure}

In our work, we empirically estimate the number of Trotter steps required to achieve convergence in the QPE algorithm, ensuring a sufficiently small error in the trotterized expansion. We plot the most likely estimated eigenvalue obtained using QPE for various combinations of ancilla qubits and Trotter steps for the (2/2) and (6/4) Hamiltonians, as shown in Fig.~SI~\ref{fig qpe convergence 2-2} and Fig.~SI~\ref{fig qpe convergence 6-4}, respectively.

Fig.~SI~\ref{fig hamiltonian 2-2} demonstrates that the choice of the number of Trotter steps, $n$, is a critical parameter to achieve high precision in the estimated phase. Examining the yellow circles, which represent the most-likely eigenvalues obtained with $n = 1$, it is clear that the results do not converge to the correct eigenvalue (indicated by the horizontal dashed line) or even reach chemical accuracy, as defined by the shaded gray area. For $n = 2$, represented by the green circles, the convergence improves significantly but still falls short of the chemical accuracy. Even with 10 ancilla qubits -- providing sufficient precision -- the estimated eigenvalues remain outside the gray area.

Finally, for $n = 3$ and $n = 4$, the estimated eigenvalues show excellent agreement with the exact eigenvalue (except for results obtained with 10 and 12 ancilla qubits). This suggests that three Trotter steps are sufficient to achieve a reasonable approximation of the electronic Hamiltonian for this system.

A different scenario is observed in Fig.~SI~\ref{fig qpe convergence 6-4}, where the energy estimates are essentially independent of the number of Trotter steps. The yellow, green, and blue circles, corresponding to the $n = 1$, $n = 2$, and $n = 3$ Trotter steps, respectively, overlap perfectly. As a result, the estimated results for $n = 2$ and $n = 3$ are shown only up to 10 ancilla.

In this case, achieving the required phase precision would necessitate 13 ancilla qubits. However, due to simulation limitations, this data point could not be obtained. Nonetheless, with 12 ancilla qubits, the estimated eigenvalues already fall very close to the shaded area.

\FloatBarrier

\section{Variational Quantum Eigensolver} \label{SI: sec2}

\begin{figure}[h]
    \centering
    \includegraphics[width=1\linewidth]{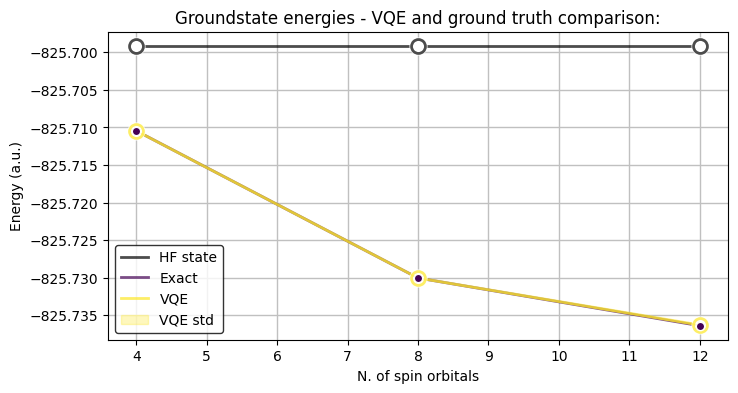}
    \caption{Ground state energies obtained with VQE (mean and standard deviation from 10 trials) and explicit diagonalization of the Hamiltonian (``Exact'' label). As a reference Hartree-Fock state energies was added (``HF''  label). Experiments were run for three active spaces (2/2) - 4 qubits, (6/4) - 8 qubits and (10/6) - 12 qubits for iteration limit of 100, and the optimizer error tolerance of $10^{-5}$.}
    \label{SI_fig:methods_vqe_results}
\end{figure}

\bmhead{Variational Quantum Eigensolver results}
For three small active spaces, the convergence of VQE was checked. The mean and standard deviation from the result of 10 runs are shown at Fig.~SI~\ref{SI_fig:methods_vqe_results}. For these experiments VQE was run on noiseless, ideal simulators, i.e. where the expectation values were computed from the final state vector, not estimated from a sampled quasi-distribution. The optimization iteration limit was set to 100 and the error tolerance was set to $10^{-5}$.

\begin{figure}[h]
    \centering
    \includegraphics[width=0.95\linewidth]{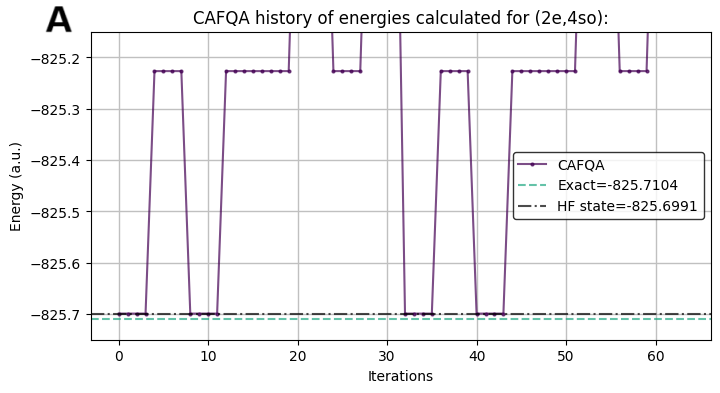}
    \includegraphics[width=0.95\linewidth]{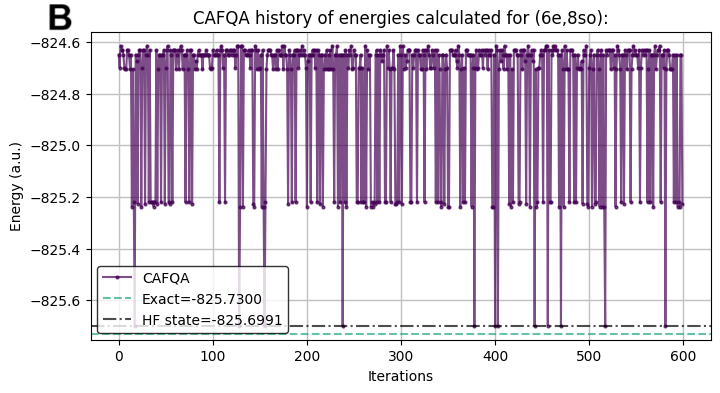}
    \caption{CAFQA warm start energies history for two active space sizes (2/2) and (6/4).}
    \label{SI_fig:CAFQA_results}
 \end{figure}

\bmhead{Clifford states warm-start for the Variational Quantum Eigensolver} Warm start is a technique for variational optimization algorithms testing particular sets of initial parameters, allowing for faster convergence of the given problems. In VQE, UCCSD and other ansaetze can be bind with a special set of parameters, that yield circuits containing only Clifford gates.~\cite{grier2022classification, vlasov1999quantum}. These quantum circuits can be efficiently simulated on classical computers using so-called stabilizer simulators.~\cite{aaronson2004improved}
In this paper, we have used this set of parameters to run the initial few iterations of VQE on stabilizer simulators, to see if we can converge towards the energies lower than by using the HF state, which was the initial starting point for the default VQE implementation (that means that initially all parameters in the UCCSD ansatz were zero). This approach, called CAFQA framework is described in this paper.~\cite{ravi2022cafqa} The results for active spaces (2/2) and (6/4) are shown in Fig.~SI~\ref{SI_fig:CAFQA_results}. No improvement for two smallest active space sizes was observed over Hartree-Fock state initial set of parameters.

\begin{figure}[h]
    \centering
    \includegraphics[width=\linewidth]{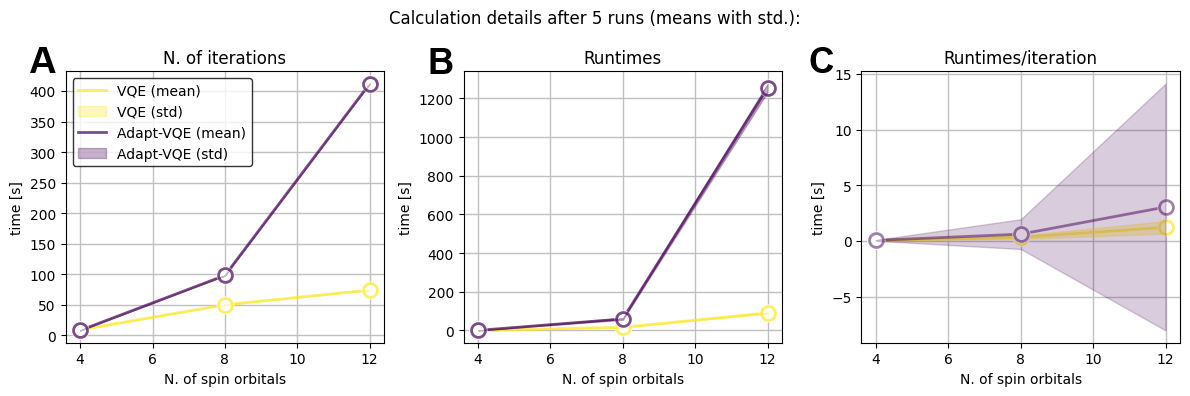}
    \caption{VQE runs data for all Adapt-VQE and default VQE implementation in Qiskit for active space sizes from (2/2), (6/4) and (10/6).}
    \label{SI_fig:Adapt_VQE_comparison}
 \end{figure}

\bmhead{Adapth-VQE results for the Variational Quantum Eigensolver} 
Adapt-VQE is a variation of VQE that aims to improve its convergence by iteratively including more parameterized terms in the ansatz, starting from a small number of terms. Per each sub-run, an ansatz with the initially included terms is run with the usual VQE optimization until it converges, and then another set of terms is included. For our case, the default VQE is systematically faster, with a smaller number of iterations per each run, than Adapt-VQE (Fig.~SI~\ref{SI_fig:Adapt_VQE_comparison}).

\FloatBarrier

\begin{figure}[h]
    \centering
    \includegraphics[width=0.81\linewidth]{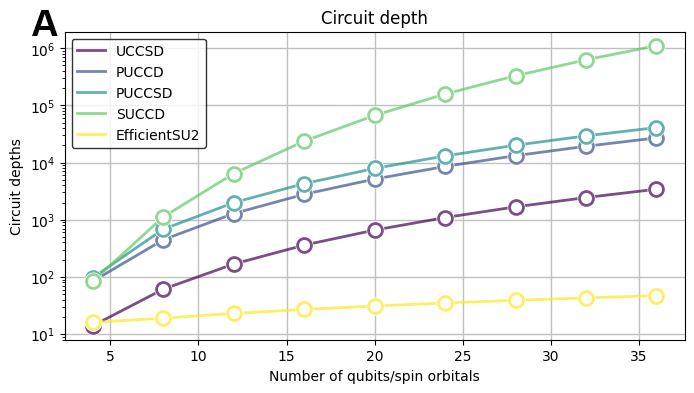}
    \includegraphics[width=0.81\linewidth]{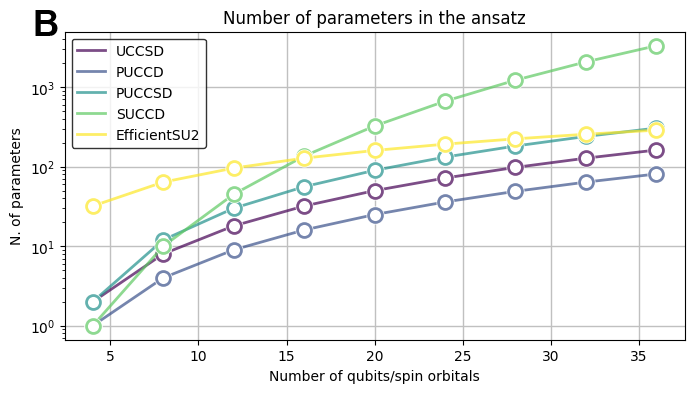}
    \includegraphics[width=0.81\linewidth]{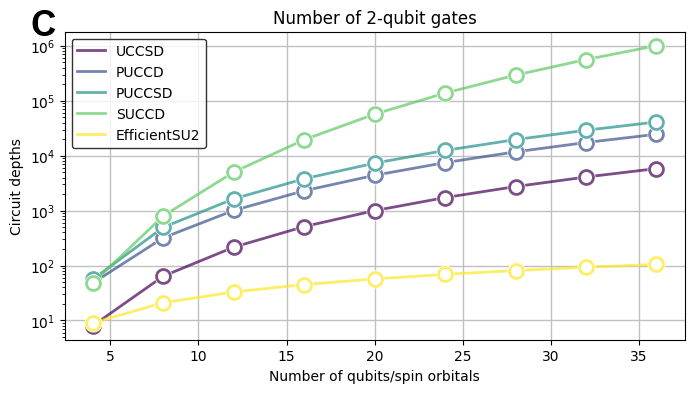}
    \caption{Quantum circuit metrics for all available ansaetze in Qiskit synthesis for active space sizes from (2/2) to (34/18) with iterative increase of number of electrons and number of spin orbitals per 4. Metrics respectively from the top: (A) circuit depth, (B) number of parameters in the ansatz and (C) number of 2-qubit gates.}
    \label{SI_fig:VQE_ans_comparison}
 \end{figure}

\bmhead{Ansatze comparison for the Variational Quantum Eigensolver} 
For systematic comparison of the circuits produced by different ansatze and their scaling towards the large active spaces, a few types of ansatze were synthesized with Qiskit and transpiled for simulator backends for Clifford$+\text{R}_z$ gates with all-to-all qubit connectivity. The results of the key metrics for these are plotted in the Fig.~SI~\ref{SI_fig:VQE_ans_comparison} for comparison. 

\FloatBarrier

The UCCSD ansatz yields the best circuit depth scaling and the second best parameter count scaling to the chemistry-informed ansatze, hence it was chosen for the computations. 
\FloatBarrier

\bmhead{UCCSD ansatz terms contribution for the Variational Quantum Eigensolver} 
 
\begin{figure}[h]
    \centering
    \includegraphics[width=0.8\linewidth]{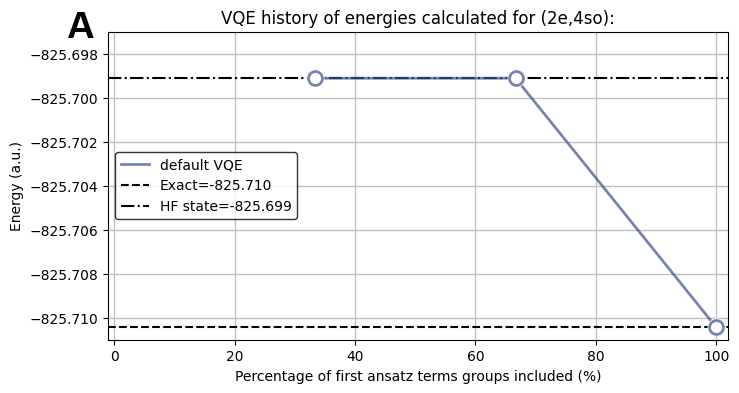}
    \includegraphics[width=0.8\linewidth]{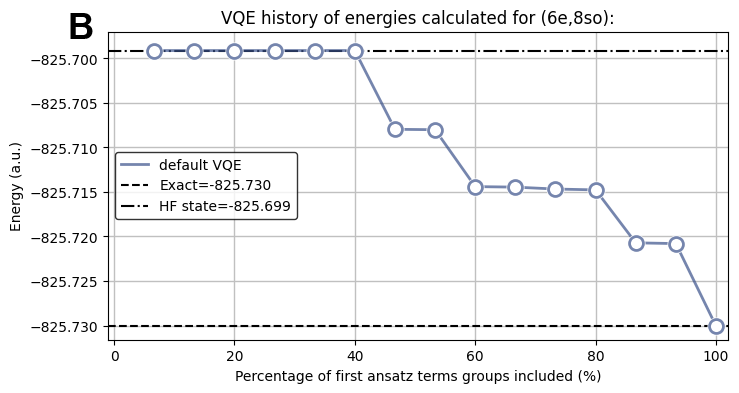}
    \includegraphics[width=0.8\linewidth]{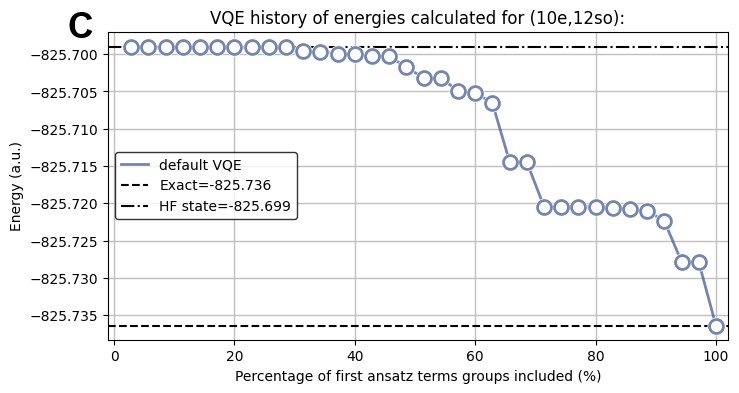}
    \caption{Terms contribution to the exact ground state energy convergence history for three active space sizes (2/2), (6/4) and (10/6).}
    \label{SI_fig:VQE_terms_contribution_results}
 \end{figure}

The UCCSD ansatz, like any other ansatz, might contain the groups of gates which are redundant to the convergence from HF state towards the true ground state configuration. If the ansatz might converge towards the ground state energy configuration with just some minority of terms, there might be a potential for excluding some terms in the ansatz, hence reducing the circuit depth and its number of parameters. To check if this is the case for UCCSD in our problem, VQE was run iteratively for different numbers of groups of the ansatz' terms sharing the same parameters, starting from single group (meaning the single parameter) and adding one group per each iteration until the VQE runs for the whole ansatz. The energies obtained for those runs are shown at Fig.~SI~\ref{SI_fig:VQE_terms_contribution_results}. Each group of terms irregularly decrease the converged energy, but for two largest tested active spaces (after treating (2/2) as an outlier), most of the terms contribute to the energy convergence towards exact ground state energy and all the terms are needed to converge towards exact solution. 
 
This shows little room for improvement when it comes to excluding terms from the ansatz. Potentially, excluding first terms could be tested, if they contribute or not, but even if the terms alone do not allow for the convergence towards lower energies, they may change the initial state, so that it becomes a proxy towards exact ground state converged with addition of latter terms, in particular cases making them necessary for the ansatz to converge.

\FloatBarrier

\bmhead{Minimum shot count estimation for the variational quantum eigensolver} 
\textbf{Single observable case.} For setting up the details of the VQE experiments, one needs to decide on the number of shots for QPU runs to estimate the energy and evaluate the gradients for the optimizer. This number decides about the accuracy of the whole algorithm (particularly in low-quantum-noise scenario for quantum resource estimation), by reducing the statistical noise coming from stochastic nature of the projective measurement of qubits. 
In this paper, the minimum number of shots will be numerically bounded by the maximum standard deviation (std) of the energy estimated by the VQE. 

We will start the bounding analysis from the definition of the intrinsic variance (for a single experiment) of the exemplary observable $\sigma^2(\hat{A})$:
$$
\sigma^2( \langle \hat{A} \rangle) =  \langle \hat{A}^2 \rangle - \langle \hat{A} \rangle ^2.
$$
Since the observable $\hat{A}$ is a tensor product of Pauli matrices $\hat{A} = \hat{P}_1 \otimes ... \otimes \hat{P}_k \otimes ... \otimes  \hat{P}_K$, where $\hat{P_k} \in \{\hat{I}, \hat{X}, \hat{Y}, \hat{Z}\}$, and each Pauli matrix is involutory ($\hat{P}_k^2=\hat{I}$), $\hat{A}$ is also involutory:
$$
\hat{A}^2 = (\hat{P}_1 \otimes ... \otimes  \hat{P}_K)^2 = \hat{P}_1^2 \otimes ... \otimes  \hat{P}_K^2 = I^{\otimes K}.
$$
This changes the intrinsic variance formula to:
$$
\sigma^2(\langle \hat{A} \rangle) =  \langle (\hat{I}^{\otimes K})^2 \rangle - \langle \hat{A} \rangle ^2 = \mathbf{1} - \langle \hat{A} \rangle ^2.
$$
Moving on into the variance of series of experiments:
$$
\mathrm{Var}(\langle \hat{A} \rangle) = \frac{\sigma^2(\langle \hat{A} \rangle)}{N}
$$
where $N$ denotes number of shots. The standard deviation is is defined as the square root of the variance, $\mathrm{Std}(\langle \hat{A} \rangle)\equiv\sqrt{\mathrm{Var}(\langle \hat{A} \rangle)}$. Then
$$
\mathrm{Std}(\langle \hat{A} \rangle) = \sqrt{\frac{\sigma^2(\langle \hat{A} \rangle)}{N}}.
$$
This way we can bound the minimum number of shots, so that the standard deviation of the observable will be smaller than defined threshold $\epsilon$:
$$
\mathrm{Std}(\langle \hat{A} \rangle) = \sqrt{\frac{\sigma^2(\langle \hat{A} \rangle)}{N}} \leq \epsilon,
$$
and so
$$
N \geq \frac{\mathbf{1} - \langle \hat{A} \rangle ^2}{\epsilon^2}.
$$
$\langle \hat{A} \rangle ^2$ takes values from 0 to 1, so in the most pessimistic scenario for $N$ to be the largest: 
$$
N \geq \frac{1}{\epsilon^2}.
$$

\textbf{Many observables case.} 
The next step is generalizing, so that the threshold $\epsilon$ bound the std of the energy $E$ measured as a weighted sum of observable expectation values $\mathbf{E}[E]=\sum_i c_i \langle \hat{A}_i \rangle$, not just a single observable expectation value. Assuming that the energy measurements for all $m$ observables are independent, with the Bienaymé's formula \cite{Wasserman2004}:
$$
\mathrm{Var}(E) = \mathrm{Var}(\sum_i c_i \langle \hat{A}_i \rangle) = \sum_i \mathrm{Var}(c_i \langle \hat{A}_i \rangle),
$$
and properties of linear scaling of random variables \cite{Wasserman2004}, we get the formula for variance $E$:
$$
 \sum_i \mathrm{Var}(c_i \langle \hat{A}_i \rangle) =  \sum_i c_i^2 \cdot \mathrm{Var}(\langle \hat{A}_i \rangle).
$$
Let us unwrap the variance of energy even further, deciding that per each observable the measurement is repeated $n_k$ times:
$$
\mathrm{Var}(E) = \sum_i^m c_i^2 \cdot \mathrm{Var}(\langle \hat{A}_i \rangle) = \sum_i^m c_i^2 \cdot \frac{{\sigma^2}(\langle \hat{A}_i \rangle)}{n_i} = \sum_i^m c_i^2 \cdot \frac{\mathbf{1} - \langle \hat{A}_i \rangle}{n_i}
$$
The worst-case scenario, which yields the largest variance of $E$, occurs when all the absolute values of expectation values are $1$. Then:
$$
\mathrm{Var}(E) = \sum_i^m   \frac{c_i^2}{n_i}
$$
and to bound the standard deviation by $\epsilon$:
$$
\mathrm{Std}(E) = \sqrt{\sum_i^m  \frac{c_i^2}{n_i}} \leq \epsilon.
$$
If we take the same number of shots per each observable ($n_k=n$ for all $k$), the minimum number of shots $N$ for the desired maximum std. of the measured energy is:
$$
N = n \cdot m \geq \frac{m}{\epsilon ^2} \sum_i^m c_i^2.
$$

\textbf{Optimization via simultaneous measurements.}  
It is well-known that commuting observables may be measured simultaneously \cite{Patel2025} . This fact may be used to decrease the number of measurements required in VQE by partitioning Hamiltonian terms into commuting subgroups. As before, let us denote the Hamiltonian in terms of its constituent terms as $E = \sum_{i}{c_i A_i}$. Let $\{G_1, \dots, G_l\}$ be a partition of the indices of the observables $\{A_i\}$, so that
$$
    E = \sum_{j=1}^{l} {\sum_{i \in G_j} {c_i A_i} }.
$$
Then, taking $n_j$ shots per group of observables $G_j$,
\begin{align*}
    \mathrm{Var}(E) &= \sum_{j=1}^{l} {\sum_{i \in G_j} {c_i^2 \mathrm{Var}(A_i)} }\\
    &\leq \sum_{j=1}^{l} { | G_j | \max_{i \in G_j}(c_i^2 \mathrm{Var}(A_i))} \\
    &\leq \sum_{j=1}^{l} { \frac{| G_j |}{N_j} \max_{i \in G_j}(c_i^2)}.
\end{align*}

To achieve an overall target variance of $\epsilon^2$, the following calculation shows that it suffices to take $n_j = |G_j| l \max_{i \in G_j}(c_i^2) / \epsilon^2$ shots per observable group $G_j$.
\begin{align*}
    \mathrm{Var}(E) &\leq \sum_{j=1}^{l} { \frac{| G_j |}{N_j} \max_{i \in G_j}(c_i^2)} \\
    &= \sum_{j=1}^{l} { \frac{| G_j |}{|G_j| l \max_{i \in G_j}(c_i^2) / \epsilon^2} \max_{i \in G_j}(c_i^2)} \\
    &= \epsilon^2 \sum_{j=1}^{l}{\frac{1}{l}} \\
    &= \epsilon^2.
\end{align*}

\begin{figure}[h]
    \centering
    \includegraphics[width=0.9\linewidth]{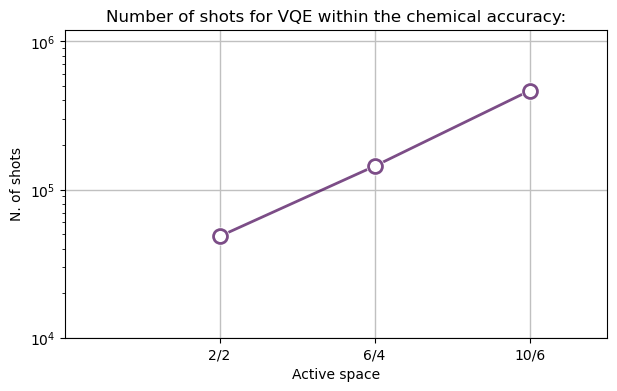}
    \caption{Minimum number of shots bounding standard deviation of the estimated energy due to statistical error to threshold $\epsilon=1.6$ mHa for the ground state energy estimation with VQE. Values were calculated based on Hamiltonians of (2/2), (6/4) and (10/6) active spaces.}
    \label{SI_fig:VQE_n_of_shots}
 \end{figure}

In this paper, the standard deviation of the energy is bounded by the threshold derived from the concept of chemical precision, as $\epsilon=1.6$ mHa. Exemplary values of the minimum number of shots bounded this way are displayed at Fig.~SI~\ref{SI_fig:VQE_n_of_shots}. 

\FloatBarrier

\section{Parameters for the Resource Estimation} \label{SI: sec3}
For the quantum resource estimation, we utilized the resource estimation tool provided by Microsoft Azure.~\cite{javadi2024quantum} Resource estimation for both VQE and QPE was conducted using the default parameters based on the physical qubit specification \texttt{qubit\_gate\_ns\_e4}. These parameters correspond to future superconducting qubit architectures operating in the nanosecond regime.
For this analysis, we sticked to all-to-all connectivity, since the real-backend connectivity overhead may be easily countered with more effective circuit synthesis~\cite{ostmeyer2023optimised, mukhopadhyay2023synthesizing, van2020circuit}, decomposition~\cite{bravyi2017tapering, harrow2024optimal} and transpilation methods~\cite{setia2018bravyi, miller2023bonsai, chiew2023discovering}. 

Specifically, gate and measurement operation times were assumed to be 50 ns and 100 ns, respectively. Idle, single-qubit, two-qubit, t-gate and measurement error rates were set to $10^{-4}$, which represents an optimistic target for such quantum computing architectures.
\newpage

\section{DFT Calculations} \label{SI: sec4}

\begin{table}[ht]
\centering
\caption{Charge on each oxygen atom, decomposed into orbitals listed in the first column. The second column shows values for a free O$_2$ molecule, while the next five columns correspond to the five geometries considered in NEB calculations. In the TS$_{\text{diff}}$ configuration, the charge differs between the two oxygen atoms, and both values are indicated. The total charge values highlight the first electron transfer in the Reactant and the second transfer along the reaction pathway, with \textit{cis} and \textit{trans} configurations exhibiting the strongest charge transfer.}
\begin{tabular}{@{}lllllll@{}}
\toprule
\textbf{Orbital} & \textbf{O$_2$} & \textbf{Reactant} & \textbf{TS$_{\text{diss}}$} & \textbf{\textit{cis}} & \textbf{TS$_{\text{diff}}$} & \textbf{\textit{trans}} \\ 
\midrule
$2S$ & 1.925 & 1.851 & 1.909 & 1.882 & 1.907; 1.871 & 1.858 \\ 
$2P_z$ & 1.497 & 1.681 & 1.627 & 1.630 & 1.502; 1.684 & 1.697 \\ 
$2P_x$ & 1.497 & 1.820 & 1.772 & 1.630 & 1.648; 1.658 & 1.678 \\ 
$2P_y$ & 1.060 & 1.079 & 1.297 & 1.730 & 1.754; 1.658 & 1.678 \\ 
\midrule
Total & 5.979 & 6.431 & 6.606 & 6.891 & 6.811; 6.871 & 6.912 \\ 
\botrule
\end{tabular}

\label{tab:orbital_values}
\end{table}

In addition to the explanations provided in the main text, Table~\ref{tab:orbital_values} presents a
detailed decomposition of the charge transfer along the reaction path by analyzing the charge values
for each oxygen atom in the five geometries considered in the NEB calculations.

The first column lists the oxygen orbitals, while the second column provides the corresponding charge for a free O$_2$ molecule. The subsequent columns display the charge values for five geometries considered: Reactant, TS$_{\text{diss}}$ (transition state for the dissociation of the oxygen atom), \textit{cis} 
configuration, TS$_{\text{diff}}$ (transition state for the diffusion of the oxygen atoms on the Cu surface), 
and \textit{trans} configuration.

The charge values confirm that, in the Reactant geometry, where O$_2$ is deposited on the Cu 
surface, an initial charge transfer occurs, increasing the total charge from 5.979 (isolated O$_2$ 
molecule) to 6.431. This additional charge mainly populates the $2P_z$ and $2P_x$ orbitals, indicating a strong 
interaction with the substrate. As the reaction progresses, further charge redistribution takes 
place as mentioned in the article, particularly between the Reactant and \textit{trans} configurations.
The \textit{cis} and \textit{trans} geometries 
exhibit the highest total charge values (6.891 and 6.912, respectively), indicating that oxygen 
atoms undergo a stronger electronic interaction with the Cu surface in these states. In these 
configurations, the $2P_y$ orbital shows a notable charge increase, suggesting that lateral 
interactions with the Cu atoms influence the electronic structure of oxygen.

A particularly interesting feature appears in the TS$_{\text{diff}}$ configuration, which is the only case 
where the two oxygen atoms exhibit different charge values. In all other geometries, both atoms 
remain in symmetric positions (either in hollow or bridge sites on the Cu surface), leading to 
nearly identical charge distributions. However, in TS$_{\text{diff}}$, one oxygen atom is positioned in a 
hollow site while the other is in a bridge position, causing an asymmetry in charge distribution. 
This is particularly evident in the $2P_z$ and $2P_y$ orbitals, where charge values differ significantly between the two oxygen atoms.

Overall, this decomposition of the charge transfer provides a more detailed understanding of how the 
electronic structure evolves along the reaction path. The strongest charge transfer occurs as the 
system progresses from the Reactant to the final \textit{trans} configuration, with notable charge 
accumulation in specific orbitals.

\begin{figure}[ht]
\centering
\includegraphics[width=0.95\textwidth]{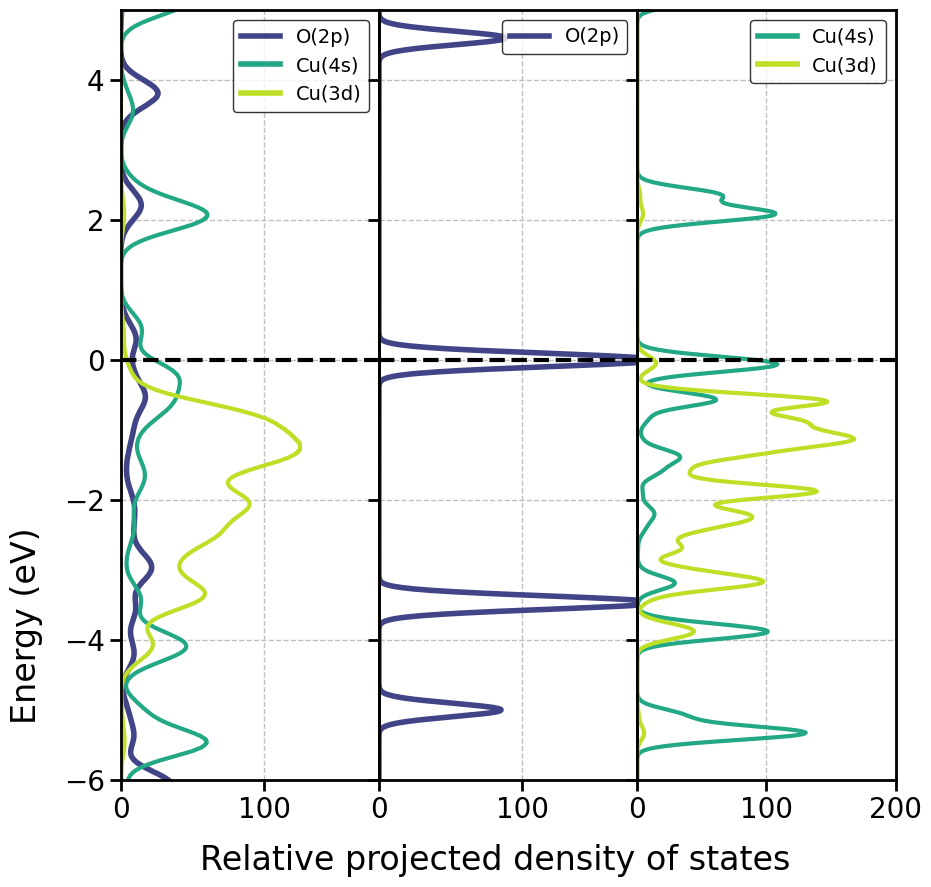}
\caption{Projected density of states (PDOS) for different atomic configurations: the first panel represents the PDOS for the oxygen molecule on top of the Cu slab in the Reactant geometry as in the main text using the same colors scheme. The last two panels correspond to a Cu slab system with an \ch{O2} molecule placed far away. The energy (in eV) is shown on the vertical axis, and the relative projected density of states is plotted on the horizontal axis. The dashed line indicates the Fermi level.} \label{fig:pdos_free}
\end{figure}

To gain insight into the electronic interactions between the O\textsubscript{2} molecule and the Cu surface, we analyze the projected density of states (PDOS) for different atomic configurations, as shown in Fig.~SI~\ref{fig:pdos_free}.

The first panel presents the PDOS for the 2$p$ orbital of oxygen and the 4$s$ and 3$d$ orbitals of Cu when the O\textsubscript{2} molecule is adsorbed on top of the Cu surface in the reactant geometry. In contrast, the second and third panels correspond to a Cu slab system with an O\textsubscript{2} molecule placed far away from the surface. The second panel displays the PDOS of the 2$p$ orbital of the isolated O\textsubscript{2} molecule, while the third panel shows the PDOS of the 4$s$ and 3$d$ orbitals of Cu in the absence of direct interaction with O\textsubscript{2}.

By comparing these panels, we can clearly identify the splitting of the $\sigma^*$ and $\pi^*$ molecular levels upon adsorption. This observation aligns with the discussion in the main text, where we describe the hybridization of these molecular states with the Cu 4$s$ and 3$d$ bands. Such hybridization is a key factor in understanding the electronic structure modifications induced by adsorption and their implications for the catalytic activity of the Cu surface.

\newpage

\section{Active Space} \label{SI: sec5}
\begin{figure}[ht]
\centering
\includegraphics[width=0.5\textwidth]{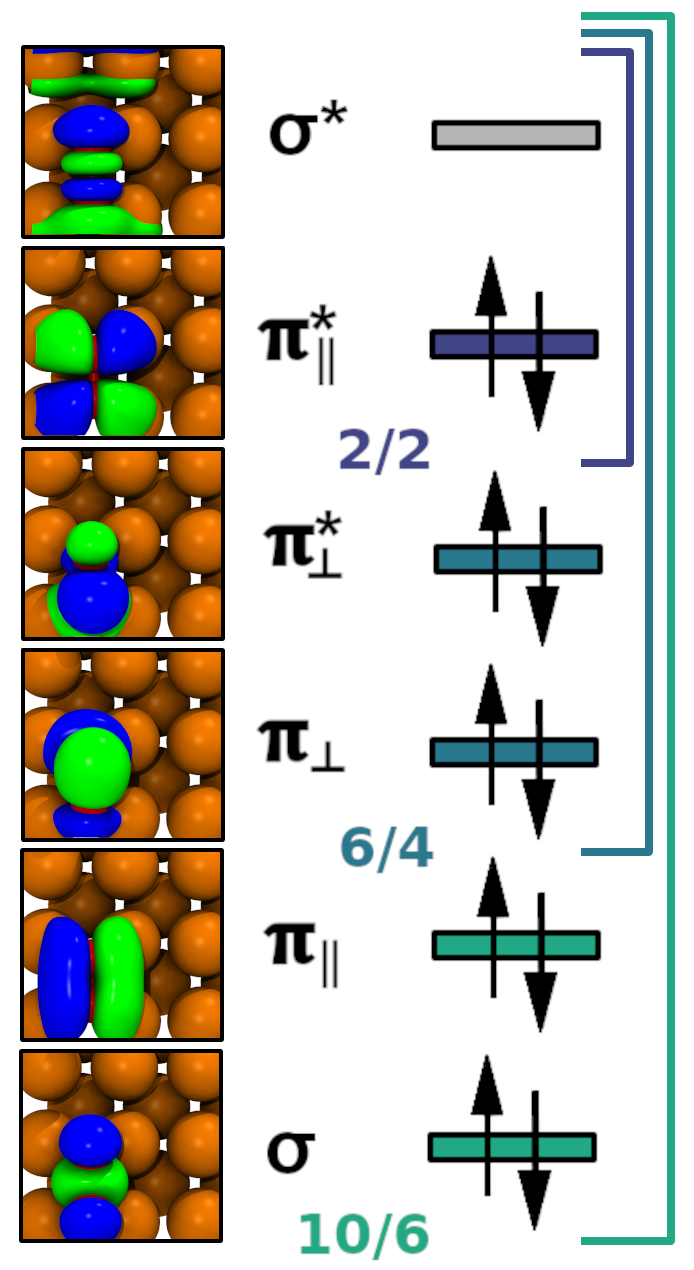}
\caption{Graphical representation of the active space orbitals extracted from the AVAS calculations used for the three active spaces: (2/2), (6/4) and (10/6). All of them contain a single unoccupied orbital in the HF reference, that is the $\sigma^*$, and an increasing number of occupied orbital: $\pi^*_{\parallel}$, $\pi^*_{\perp}$, $\pi_{\perp}$, $\pi_{\parallel}$ and $\sigma$.} \label{fig:active space}
\end{figure}

\bibliography{si}